# Formalization and quantitative metrics for functional stability of edge computing systems

**Oleksii S. Bychkov**
*Taras Shevchenko National University of Kyiv, Bohdana Gavrilishina,24 str., Kyiv, Ukraine*
*Email: oleksiibychkov@knu.ua*

Abstract

Edge computing systems operate in conditions where component failures, intermittent backhaul connectivity, resource exhaustion and adversarial disturbances are part of the normal operating envelope rather than rare events. Classical reliability and fault-tolerance models, oriented toward monolithic systems and binary working/failed states, do not capture the differentiated quality requirements of services co-located on a resource-constrained edge node. This article proposes a formal framework of *functional stability* that shifts the unit of analysis from the system to the individual function. Strong and weak forms of stability are defined through a continuous quality function and per-function quality thresholds. A system of base parameters (disturbance tolerance, recovery time, degradation depth, admissible degradation) and aggregate metrics (cumulative quality, functional availability, integral stability) is introduced and connected to the formal definitions by nine theorems. The framework is instantiated for edge computing scenarios and illustrated through two analytical case studies: a UAV swarm and an edge AI inference service with cloud fallback. The proposed metrics distinguish architectural alternatives that classical availability cannot separate and expose the dependence of architectural ranking on the explicitly selected disturbance-class set.

## 1. Introduction

Edge computing has moved the locus of data processing from centralized data centres to the network periphery, where compute resources are placed close to data sources and end users [1, 2]. The architectural shift addresses bandwidth limits, end-to-end latency budgets and data sovereignty constraints that cloud-only deployments cannot satisfy [3, 4]. With this shift, however, a class of dependability problems that were peripheral in cloud settings becomes central. Edge nodes are deployed in physically exposed locations, run without on-site maintenance, draw power from constrained sources and rely on wireless or wide-area links whose availability is intermittent by design [5, 6].

Failures in edge systems are therefore not rare events to be avoided but expected modes of operation. A roadside unit may lose its uplink to the cloud during congestion peaks. A battery-powered sensor cluster may downclock its inference model when thermal limits are reached. A wearable medical device may fall back to on-device processing when its companion gateway is offline. In each case, the service does not stop; it continues with reduced quality. The system is, in a meaningful sense, still working, but classical reliability theory has no vocabulary for this state [7].

The dominant frameworks for system dependability were developed for a different operating context. The concept of *resilience*, codified in IEEE Std 24765 [8], characterizes a system as a whole and aggregates response and recovery into a single high-level property. The concept of *fault tolerance*, defined in IEC 61508 [9], partitions states into working and failed and addresses the continuation of a required function in the presence of internal faults. The concept of *dependability*, structured by ISO/IEC 25010 [10] and the taxonomy of Avizienis et al. [11], decomposes the property into availability, reliability, safety, integrity and maintainability, each evaluated by its own metric. These frameworks share three features that limit their direct use in edge contexts: a system-centric unit of analysis, a binary or ordinal evaluation scale, and an implicit treatment of degradation as something to be avoided rather than designed.

The development of the theory of functional stability is inseparable from the foundational works of Ukrainian researchers. The philosophical basis was formulated in the works of O. A. Mashkov [12, 13], who established the priority of system “survival” over the preservation of its structure. The instrumental toolkit for the practical realization of this concept, based on graph theory and structural analysis, was developed by O. V. Barabash [14, 15]. The mathematical apparatus for the optimization of stability parameters was advanced by Yu. V. Kravchenko [16, 17], and the telecommunications basis of network reliability was provided by V. V. Vyshnivskyi [18, 19].

The need for finer-grained models is recognized in the edge computing literature itself. Surveys of reliability and availability modelling in edge and fog environments [20] note that services co-located on the same node typically have heterogeneous quality requirements: a control loop tolerates milliseconds of interruption while a telemetry aggregator tolerates seconds; a safety-critical inference must run locally while a logging service can be deferred. QoS-aware provisioning frameworks for fog computing [21] and approaches based on service migration [22, 23], checkpoint–restart [24] and adaptive model selection [25] provide mechanisms for handling such heterogeneity, but they lack a unifying metric that allows architectural alternatives to be compared.

This article proposes such a metric. The concept of *functional stability* shifts the unit of analysis from the system to the individual function and replaces the binary working/failed evaluation with a continuous quality function. The two forms of the definition, strong and weak, capture two operating regimes: zero-downtime systems where no quality dip is admissible, and graceful-degradation systems where temporary quality loss is acceptable provided recovery is bounded. The framework is consistent with the standard apparatus of systems theory [26] and reliability analysis [27]; it does not replace existing concepts but provides a function-level layer that they lack.

The contributions of the article are as follows.

1. A formal definition of functional stability in strong and weak forms (Section 3.2), based on a continuous per-function quality function and a per-function quality threshold.

2. A system of base parameters and aggregate metrics (Section 3.3), each connected to the formal definitions by an explicit theorem (Section 3.4, Theorems 1–9).

3. An instantiation of the framework for edge computing (Section 4.1), with a mapping of the abstract objects (functions, disturbances, quality, recovery) to concrete edge constructs (services, link drops, latency budgets, service migration).

4. Two analytical case studies (Section 4.2 and 4.3) that show both how integral stability separates architectures with similar nominal availability and why the disturbance-class set $K_i$ must be fixed explicitly before comparing designs.

The article is organized along the IMRaD structure recommended by the journal. Section 2 reviews related work on dependability in edge and fog computing. Section 3 develops the formal framework. Section 4 presents the edge computing instantiation and the case studies. Section 5 discusses the framework and its limitations. Section 6 concludes.

# 2. Related work

## 2.1. Resilience, fault tolerance, dependability

Three concepts dominate the literature on system dependability. *Resilience* [8, 28] describes a two-phase response to disturbances: resistance and recovery. The resilience engineering paradigm formulated by Hollnagel, Woods and Leveson [28] identifies four capabilities of resilient systems: response, monitoring, anticipation and learning. Hosseini, Barker and Ramirez-Marquez [29] survey resilience metrics and observe that most classify the system on a binary or ordinal scale; Henry and Ramirez-Marquez [30] propose quantitative metrics but at the system level, without per-function granularity.

*Fault tolerance* [9, 11] addresses the continuation of a required function in the presence of internal faults. The taxonomy of Avizienis et al. [11] distinguishes faults (causes), errors (state-level manifestations) and failures (specification-level deviations). Mechanisms of fault tolerance, including redundancy, voting and recovery blocks [31], aim to prevent the cascade from fault to failure. Graceful degradation, treated by Shelton, Koopman and Nace [32] as a separate framework adjacent to fault tolerance, allows the system to shed non-critical functions to preserve critical ones.

*Dependability* [10, 11, 33] decomposes the property of a trustworthy system into five attributes: availability, reliability, safety, integrity and maintainability. The decomposition supports clear definitions but creates the inverse problem of integration: each attribute is evaluated by its own metric, and combining them into a single quantitative judgement of system fitness remains open [34].

### 2.2. Dependability mechanisms in edge and fog computing

Edge-specific fault tolerance has been studied along three axes. *Replication and redundancy* [35, 36] mirror cloud techniques but are constrained by the limited resources of edge nodes; full active replication is rarely affordable. *Service migration* [22, 23, 37] moves services between edge nodes or between edge and cloud in response to load, failure or mobility. Wang et al. [22] formalize the migration decision as a Markov decision process; Ouyang et al. [23] propose a Lyapunov-based online algorithm. *Adaptive computation* [25, 38] adjusts the workload itself, switching between full and quantized models, or between local inference and cloud offloading, when resources or links degrade.

Reviews of reliability in edge and fog environments [20] and surveys of QoS-aware fog provisioning [21], read against the broader software reliability literature [39], converge on a recurring observation: existing mechanisms provide local optimizations but lack a system-level objective that quantifies how well a particular architectural choice serves the heterogeneous quality requirements of co-located services. Service-level objectives (SLOs) [40, 41] formalize per-service requirements but stop at the boundary of an individual service; they do not aggregate to a system-level judgement that accounts for criticality weights or recovery dynamics.

### 2.3. Position of the proposed framework

The framework proposed in this article is positioned at the intersection of these strands. It does not replace resilience, fault tolerance or dependability; it provides a continuous, per-function evaluation layer that connects to SLO practice on the operational side and to formal stability concepts on the theoretical side. The detailed comparison is presented in Section 5.

## 3. Methods: the formal framework

### 3.1. System model and base notation

Formalization of the concept of functional stability is built on the standard apparatus of systems theory [26] and reliability analysis [27]. The central idea is the transition from a system-centric to a function-centric view: instead of evaluating the system as an indivisible whole, the framework considers a set of functions, each with its own quality parameters and stability requirements.

Let $S$ denote a system that realizes a finite set of functions $F = \{f_1, f_2, \dots, f_n\}$. The decomposition is consistent with the SADT/IDEF0 methodology of functional modelling [42], in which the system is represented as a hierarchy of functions linked by data and control flows. Depending on the level of abstraction, $f_i$ may correspond to a microservice, a software module or a business process. In the edge computing instantiation (Section 4), $f_i$ corresponds to a service running on an edge node, for example a local inference service, a telemetry aggregator or a control loop.

For each function $f_i$, a quality function $Q_i: \mathbb{R}_+ \to [0, Q_i^{\max}]$ is defined as a *utility-oriented* indicator: $Q_i(t)$ is constructed so that higher values denote better operation. For metrics that are intrinsically "higher is better" (throughput, inference accuracy, frame rate, control-loop frequency), $Q_i(t)$ is the raw metric. For metrics that are intrinsically "lower is better" (end-to-end latency $L_i(t)$, response time, error rate, energy per operation), a monotone-decreasing transformation is applied. A canonical choice for latency is

$$Q_i(t) = \frac{L_i^{\max}}{L_i(t)},$$

where $L_i^{\max}$ is the admissible latency budget. With this transformation, the nominal latency $L_i^{\text{nom}}$ yields $Q_i^{\text{nom}} = L_i^{\max}/L_i^{\text{nom}} \geq 1$, the budget boundary $L_i(t) = L_i^{\max}$ yields $Q_i^{\min} = 1$, and the condition "$Q_i(t) \geq Q_i^{\min}$" applies uniformly to all metric types. The continuous representation aligns with quality-of-service modelling [40].

A *minimal admissible quality threshold* $Q_i^{\min} \in (0, Q_i^{\max}]$ is associated with each function. The threshold corresponds conceptually to the service-level objective (SLO) [40, 41] used in cloud and edge operations. The use of a per-function threshold formalizes the principle of differentiated requirements: a control loop may carry a high threshold, while a logging service carries a much lower one.

The *set of admissible disturbances* $D$ contains all influences capable of degrading the operation of the system. Unlike the classical notion of fault [11], which is restricted to internal defects, the set $D$ includes both internal faults and external disturbances such as overload, network partition, cyberattack or environmental change [43]. Each disturbance $d \in D$ is characterized by a time of occurrence $t_0(d)$ and a profile of impact on the functions of the system.

The set $D$ admits a natural structure as a union of parametric classes,

$$D = \bigcup_{k \in K} D_k, \qquad D_k = \{d_k(a): a \in [0, A_k^{\max}]\},$$

where $d_k(a)$ denotes a disturbance of class $k$ with amplitude $a$. Examples of edge-relevant classes include link-loss duration, CPU overload intensity, thermal-throttling factor, battery-depletion level and peer-node churn rate. The impact of $d_k(a)$ on function $f_i$ is described by a response function $\varphi_i^{(k)}(a,t)$ such that the quality trajectory under the disturbance is $Q_i^{d_k(a)}(t) = Q_i^{\text{nom}} - \varphi_i^{(k)}(a,t)$. We assume $\varphi_i^{(k)}(0,t) = 0$ for every $t$, so amplitude $a = 0$ corresponds to the absence of degradation in class $k$ and the quality trajectory coincides with the nominal trajectory.

Throughout the framework, the response functions are assumed monotone non-decreasing in the amplitude:

$$a_1 \leq a_2 \Rightarrow \varphi_i^{(k)}(a_1,t) \leq \varphi_i^{(k)}(a_2,t) \quad \forall t, \forall i, \forall k.$$

The assumption captures the physically expected behaviour: stronger disturbances of the same class do not produce smaller degradation. It is satisfied by standard models of link loss, overload and thermal throttling, but should be checked when a class is defined by composite events.

Finally, for each function $f_i$ and disturbance $d$, the *recovery time* $T_{\mathrm{rec}}^i(d)$ is defined as the interval from the time of disturbance occurrence to the return of the function quality to the admissible level. The parameter corresponds to the Mean Time To Recovery (MTTR) [44] of classical reliability theory, but unlike the system-level MTTR, $T_{\mathrm{rec}}^i(d)$ is specified per function and per disturbance.

### 3.2. Formal definition of functional stability

Two forms of the definition are introduced, differing in the strictness of the requirement imposed on the system.

**Definition 1 (Strong form, absolute functional stability).** *A system $S$ is absolutely functionally stable with respect to a set of disturbances $D$ if*

$$\forall d \in D,\ \forall i \in \{1,\dots,n\},\ \forall t \geq t_0(d)\colon Q_i^d(t) \geq Q_i^{\min}.$$

The condition states that no disturbance from $D$ can reduce the execution quality of any function below the established threshold, not even momentarily. The requirement corresponds to the concept of zero-downtime architectures [45] used in systems with the strictest availability requirements: payment platforms, life-support medical equipment, air-traffic control. The definition formalizes the ideal of uninterrupted operation.

In practice, absolute functional stability is rarely affordable in edge contexts, where energy, compute and connectivity are all constrained. A more practical form is the weak definition.

**Definition 2 (Weak form, functional stability with recovery).** *A system $S$ is functionally stable with recovery with respect to a set of disturbances $D$ if*

$$\forall d \in D,\ \forall i\colon\ \exists T_{\mathrm{rec}}^i(d) < \infty\colon\ \forall t \geq t_0(d) + T_{\mathrm{rec}}^i(d)\colon Q_i^d(t) \geq Q_i^{\min}.$$

The weak form is consistent with the model of self-healing systems [46], in which the system detects problems and restores normal operation without external intervention. The requirement of a finite recovery time is essential: the system must be guaranteed to return to the admissible quality level rather than remain in a degraded state for an unbounded period.

The weak form does not by itself constrain the depth of temporary degradation: the quality $Q_i^d(t)$ may fall to zero on the interval $[t_0(d), t_0(d) + T_{\mathrm{rec}}^i(d))$. For applications that cannot tolerate even short complete outages of a function, additional constraints on the minimal quality during recovery are required; these are introduced through the degradation-depth parameter below.

### 3.3. Parameters of functional stability

The formal definitions of strong and weak functional stability are qualitative: they fix the conditions under which a system is functionally stable but provide no apparatus for comparison between systems. The transition to quantitative analysis requires a system of parameters, each connected to a component of the formal definition.

**Definition 3 (Disturbance tolerance).** *The disturbance tolerance of function $f_i$ with respect to disturbance class $k$ is*

$$R_d^{i,k} = \sup\{a \in [0, A_k^{\max}]:\ \forall t \geq t_0:\ Q_i^{d_k(a)}(t) \geq Q_i^{\min}\}.$$

Under the monotonicity assumption introduced in Section 3.1, every amplitude $a < R_d^{i,k}$ is admissible: if the function survives amplitude $a^*$, it survives every smaller amplitude of the same class. With the continuity assumption used in Section 3.4, the boundary amplitude $a = R_d^{i,k}$ is admissible as well. A system with high values of $R_d^{i,k}$ across all functions and classes tolerates more intense disturbances without violation of the strong stability condition. The parameter quantifies the “margin” of each function within each disturbance class.

**Definition 4 (Recovery time).** *The recovery time of function $f_i$ under disturbance $d$ is*

$$T_{\text{rec}}^i(d) = \inf\{\tau \geq 0:\ \forall t \geq t_0(d) + \tau:\ Q_i^d(t) \geq Q_i^{\min}\}.$$

The smaller the value of $T_{\text{rec}}^i(d)$, the faster the system returns to the admissible regime after disturbance $d$. The infimum is taken over $\tau \geq 0$ rather than $\tau > 0$, so that $T_{\text{rec}}^i(d) = 0$ is admissible and corresponds to a disturbance that never drives $f_i$ below the threshold. In the boundary case where $T_{\text{rec}}^i(d) = 0$ for all $d \in D$ and all $i$, the weak form coincides with the strong form, establishing a natural hierarchy between the two definitions.

Operationally, two related but distinct times must be separated: the time of return to the admissible band (the recovery time of Definition 4) and the time of return to the nominal operating point. The latter is needed for case studies of architectures that reconfigure on disturbance without crossing the threshold, and is introduced as a separate parameter.

**Definition 4a (Nominal-restoration time).** *The nominal-restoration time of function $f_i$ under disturbance $d$ is*

$$T_{\text{nom}}^i(d) = \inf\{\tau \geq 0:\ \forall t \geq t_0(d) + \tau:\ |Q_i^d(t) - Q_i^{\text{nom}}| \leq \varepsilon_i\},$$

*where $\varepsilon_i$ is a small operational tolerance around the nominal value.*

Provided the nominal tolerance band is contained in the admissible band, i.e. $Q_i^{\text{nom}} - \varepsilon_i \geq Q_i^{\min}$, one has $T_{\text{rec}}^i(d) \leq T_{\text{nom}}^i(d)$. Equality can occur in several cases, including the no-disturbance trajectory (both equal zero) and architectures in which the system returns to the admissible band and to the nominal band simultaneously; in general, $T_{\text{nom}}^{i}$ is a stricter return-to-normal indicator than $T_{\text{rec}}^{i}$. The distinction matters in edge architectures with graceful reconfiguration: a service may stay above the SLO threshold throughout a disturbance but still take noticeable time to return to its design operating point. The integral stability metric (Definition 9) and Theorem 9 both use $T_{\text{rec}}^{i}$; case studies that report reconfiguration durations should make explicit which of the two times is meant.

**Definition 5 (Degradation depth).** *The degradation depth of function $f_i$ under disturbance $d$ is*

$$\Delta Q_i(d) = Q_i^{\text{nom}} - \inf_{t \geq t_0(d)} Q_i^d(t),$$

*where $Q_i^{\text{nom}}$ is the nominal quality of function $f_i$ in the absence of disturbances and $Q_i^d(t)$ denotes the quality trajectory under disturbance $d$.*

The infimum is taken over the full post-disturbance trajectory $t \geq t_0(d)$ rather than over the recovery interval alone, so the parameter remains well-defined when $T_{\text{rec}}^{i}(d) = 0$ and when the system continues to oscillate after nominal recovery is first achieved. A system that satisfies the weak form but exhibits $\Delta Q_i(d) \approx Q_i^{\text{nom}}$ (almost complete loss of function at some point) may be unacceptable for many edge applications, so the parameter allows systems with the same recovery time but different "depth of dip" to be distinguished.

**Definition 6 (Admissible degradation).** *The admissible degradation of function $f_i$ is*

$$\delta_i = Q_i^{\text{nom}} - Q_i^{\text{min}}.$$

The parameter defines the admissible range of quality variation within which the system remains in the region of strong functional stability. The relationship between strong stability and admissible degradation is expressed by the equivalence

$$\text{Strong stability for } (f_i, d) \Leftrightarrow \Delta Q_i(d) \leq \delta_i.$$

*3.3.1. Aggregate metrics*

The base parameters describe the behaviour of a single function under a single disturbance. For practical use, aggregate indicators are needed that allow the system to be evaluated as a whole.

**Definition 7 (Cumulative quality).** *The cumulative quality of function $f_i$ over the time horizon $[0, T]$ is*

$$\mathcal{Q}_i(T) = \int_0^T Q_i\,(t)\,dt.$$

The indicator measures the total "useful work" performed by the function over the period. For a function operating at nominal quality without disturbances, the cumulative quality is $Q_i^{\text{nom}} \cdot T$. Because the quality may transiently exceed the nominal level (for example, latency briefly better than the design point), the efficiency coefficient is defined with a cap at the nominal value:

$$\eta_i(T) = \frac{1}{T}\int_0^T \min\left(1, \frac{Q_i(t)}{Q_i^{\text{nom}}}\right) dt \in [0,1].$$

The cap ensures that better-than-nominal operation does not compensate for periods of degradation, which is the operationally meaningful interpretation: an extra millisecond shaved off a latency budget that already meets the SLO does not offset a later interval in which the SLO is missed.

**Definition 8 (Functional availability).** *The functional availability of function $f_i$ over the time horizon $[0, T]$ is*

$$A_i(T) = \frac{1}{T}\int_0^T \mathbf{1}\,[Q_i(t) \geq Q_i^{\min}]\, dt,$$

*where $\mathbf{1}[\cdot]$ is the indicator function.*

For stationary processes, as $T \to \infty$, the indicator converges to the classical availability coefficient

$$A_i = \lim_{T\to\infty} A_i(T) = \frac{\text{MTBF}_i}{\text{MTBF}_i + \overline{T}_{\text{rec}}^i},$$

where $\text{MTBF}_i$ is the mean time between violations of the condition $Q_i(t) \geq Q_i^{\min}$ and $\overline{T}_{\text{rec}}^i$ is the mean recovery time. The link to the formal definitions is one-directional and is formalized in Theorem 7: the strong form for $f_i$ over $D$ implies $A_i = 1$ on every realized trajectory, but the converse holds only under a coverage assumption on the observation horizon.

**Definition 9 (Integral stability).** *The integral stability of system $S$ with respect to a set of disturbances $D$ is*

$$\Psi(S, D) = \sum_{i=1}^{n} w_i \cdot \hat{R}_i \cdot \frac{T_i^{\max}}{T_i^{\max} + \overline{T}_{\text{rec}}^i},$$

*where the weights $w_i \geq 0$, $\sum_i w_i = 1$, reflect the relative criticality of the functions; $K_i \subseteq K$ is the set of disturbance classes relevant to function $f_i$; $\hat{R}_i$ is the capped normalized disturbance tolerance*

$$\hat{R}_i = \min_{k\in K_i} \min\left(1, \frac{R_d^{i,k}}{R_{\text{ref}}^{i,k}}\right) \in [0,1],$$

*with $R_{\text{ref}}^{i,k}$ a class-specific reference amplitude derived from the operating envelope (the maximum amplitude observed or anticipated in class $k$); $T_i^{\max}$ is the maximum admissible recovery time for function $f_i$, derived from the service-level objective on time-to-recovery; and $\overline{T}_{\text{rec}}^i$ is the conditional mean recovery time given that a threshold violation has occurred,*

$$\overline{T}_{\text{rec}}^i = \mathbb{E}\left[T_{\text{rec}}^i(d) \,\middle|\, \exists t: Q_i^d(t) < Q_i^{\min}\right],$$

*where the expectation is taken under the empirical or design distribution of disturbances in $D$. By convention, if the conditioning event has zero probability (no*

*disturbance crosses the threshold for $f_i$), then $\overline{T}^{i}_{\mathrm{rec}} = 0$ and the recovery factor takes the value* 1.

The form combines two aspects of stability: normalized tolerance to disturbances ($\hat{R}_i$) and speed of recovery relative to the admissible budget ($T_i^{\max}/(T_i^{\max} + \overline{T}^{i}_{\mathrm{rec}})$). Both factors are bounded in [0,1], so $\Psi(S, D) \in [0,1]$. The metric attains $\Psi = 1$ only when all weighted functions tolerate the full specified design envelope within the declared $K_i$ and reference envelopes $R^{i,k}_{\mathrm{ref}}$, and recover instantaneously ($\overline{T}^{i}_{\mathrm{rec}} = 0$). If a weighted function fails to recover ($\overline{T}^{i}_{\mathrm{rec}} \to \infty$), its contribution tends to zero; $\Psi$ tends to zero only when all weighted contributions collapse simultaneously. The boundedness eliminates the dominance pathology of unbounded metrics, in which a single function with near-zero recovery time can mask poor performance elsewhere. Section 3.4 establishes the connection between $\Psi$, functional availability and the disturbance dynamics through the two-state failure–repair model (Theorem 9).

Table 1 systematizes the introduced parameters.

| Parameter | Notation | Link to formal definition |
| --- | --- | --- |
| Disturbance tolerance | $R^{i,k}_d$ | Per-class margin; under monotonicity, $a < R^{i,k}_d$ implies $Q_i^{d_k(a)}(t) \geq Q_i^{\min}$; the boundary $a = R^{i,k}_d$ is admissible under continuity |
| Recovery time | $T^{i}_{\mathrm{rec}}(d)$ | Central parameter of the weak form; finiteness is the defining requirement; $T^{i}_{\mathrm{rec}} = 0$ reduces the weak form to the strong form |
| Nominal-restoration time | $T^{i}_{\mathrm{nom}}(d)$ | Time of return to nominal operating point; complementary QoS indicator; $T^{i}_{\mathrm{rec}} \leq T^{i}_{\mathrm{nom}}$ |
| Degradation depth | $\Delta Q_i(d)$ | Worst quality dip over the post-disturbance trajectory; $\Delta Q_i(d) \leq \delta_i$ is equivalent to the strong form |
| Admissible degradation | $\delta_i$ | Buffer zone for the strong form: $\delta_i = Q_i^{\mathrm{nom}} - Q_i^{\min}$ |
| Cumulative quality | $\mathcal{Q}_i(T)$ | Integral $\int_0^T Q_i(t)\, dt$; total “useful work” of the function |
| Functional availability | $A_i$ | Fraction of time with $Q_i \geq Q_i^{\min}$; strong form implies $A_i = 1$ (one direction only, see Theorem 7) |

| Parameter | Notation | Link to formal definition |
|---|---|---|
| Integral stability | $\Psi(S, D)$ | Bounded aggregate: $\Psi = \sum_i w_i \cdot \hat{R}_i \cdot T_i^{\max}/(T_i^{\max} + \overline{T}^i_{\text{rec}}) \in [0,1]$ |

*Table 1. Parameters of functional stability and their link to the formal definitions*

3.4. Theorems linking the parameters to the definitions

*3.4.1. Parametric characterization of the strong form*

For the proofs in this subsection it is assumed, in addition to the monotonicity introduced in Section 3.1, that for each $i, k, t$ the response function $\varphi_i^{(k)}(a, t)$ is continuous in $a$. The continuity, together with monotonicity, makes the set $\{a \in [0, A_k^{\max}]: \forall t: \varphi_i^{(k)}(a, t) \le \delta_i\}$ closed and of the form $[0, R_d^{i,k}]$, so that the supremum in Definition 3 is in fact a maximum and the boundary amplitude $R_d^{i,k}$ itself is admissible. The assumption is satisfied by standard models of link loss, overload, thermal throttling and similar physical disturbances; piecewise-continuous response functions can be handled by splitting the amplitude domain into sub-classes on which continuity holds.

**Theorem 1 (Criterion of strong stability through disturbance tolerance).** *Under the monotonicity and continuity assumptions on the response functions, system S satisfies the strong form of functional stability for function $f_i$ with respect to class $D_k$ if and only if every disturbance $d_k(a) \in D_k$ has amplitude $a \le R_d^{i,k}$. Equivalently, $A_k^{\max} \le R_d^{i,k}$.*

*Proof.* (⇒) Suppose the strong form holds for $(f_i, D_k)$. Then for every $a \in [0, A_k^{\max}]$, $\forall t \ge t_0: Q_i^{d_k(a)}(t) \ge Q_i^{\min}$, so $a$ belongs to the set over which the supremum in Definition 3 is taken. Therefore $a \le R_d^{i,k}$ for every $d_k(a) \in D_k$.

(⇐) Suppose $a \le R_d^{i,k}$ for every $d_k(a) \in D_k$. Fix any such $a$. By Definition 3, there exists a sequence $a_m \nearrow R_d^{i,k}$ with $\forall t: Q_i^{d_k(a_m)}(t) \ge Q_i^{\min}$. By monotonicity, for the eventual $a_m \ge a$ in the sequence, $\varphi_i^{(k)}(a, t) \le \varphi_i^{(k)}(a_m, t)$, so $Q_i^{d_k(a)}(t) \ge Q_i^{d_k(a_m)}(t) \ge Q_i^{\min}$ for every $t$. For the boundary case $a = R_d^{i,k}$, continuity gives $\varphi_i^{(k)}(R_d^{i,k}, t) = \lim_m \varphi_i^{(k)}(a_m, t) \le \delta_i$, hence $Q_i^{d_k(R_d^{i,k})}(t) \ge Q_i^{\min}$ as well. The strong form holds for every $d_k(a) \in D_k$. ▫

*Remark.* Without monotonicity, the admissible-amplitude set may be non-convex and $a \le R_d^{i,k}$ does not imply admissibility of the specific amplitude $a$. Without continuity, the supremum need not be attained and the boundary point $R_d^{i,k}$ may itself fail to be admissible. In practice, both conditions should be verified when defining a disturbance class; composite events (for example, simultaneous link loss and CPU overload) typically require splitting into separate classes to preserve these properties.

**Theorem 2 (Criterion of strong stability through degradation depth).** *System S satisfies the strong form for the pair* $(f_i, d)$ *if and only if* $\Delta Q_i(d) \le \delta_i$.

*Proof.* The condition $Q_i^d(t) \ge Q_i^{\min}$ for every $t \ge t_0(d)$ is equivalent to $\inf_{t \ge t_0(d)} Q_i^d(t) \ge Q_i^{\min}$. Subtracting both sides from $Q_i^{\mathrm{nom}}$ gives $Q_i^{\mathrm{nom}} - \inf_{t \ge t_0(d)} Q_i^d(t) \le Q_i^{\mathrm{nom}} - Q_i^{\min}$, i.e., $\Delta Q_i(d) \le \delta_i$. ▫

**Corollary 1.** *The admissible degradation parameter* $\delta_i$ *defines the maximum "dip" depth compatible with the strong form of stability. A system with a higher value of* $\delta_i$ *(lower threshold* $Q_i^{\min}$ *at fixed* $Q_i^{\mathrm{nom}}$*) has a wider admissible range of quality variation.*

*3.4.2. Parametric characterization of the weak form*

For the proofs in this and the next subsection, the quality trajectories $Q_i^d(t)$ are assumed continuous in time. The assumption rules out a measure-zero pathology in which $Q_i^d$ dips below the threshold only at isolated instants: in such a case the infimum in Definition 4 would still give $T_{\mathrm{rec}}^i(d) = 0$ even though the strong form is violated at those instants. The continuity assumption is appropriate for smoothed or time-averaged quality indicators (latency from queue lengths, accuracy from windowed statistics, throughput averaged over short intervals). For inherently discrete or event-driven metrics (frame rate, telemetry update rate, container migration status, count of active replicas), the same results can be formulated for right-continuous piecewise trajectories by replacing pointwise inequalities with almost-everywhere or interval-based threshold conditions; in this generalization the infimum becomes $\inf\{\tau \ge 0: Q_i^d(t) \ge Q_i^{\min}$ for almost every $t \ge t_0(d) + \tau\}$ and gives the same numerical value under mild regularity of the disturbance process.

**Theorem 3 (Recovery time as a measure of strong-form violation).** *Let system S not satisfy the strong form for the pair* $(f_i, d)$*, i.e.,* $\Delta Q_i(d) > \delta_i$*. Then S satisfies the weak form for* $(f_i, d)$ *if and only if*

$$T_{\mathrm{rec}}^i(d) = \inf\{\tau \ge 0: \forall t \ge t_0(d) + \tau: Q_i^d(t) \ge Q_i^{\min}\} < \infty.$$

*Proof.* Direct from Definition 2 and Definition 4. ▫

**Theorem 4 (Hierarchy of stability forms).** *Under the continuity assumption on* $Q_i^d$*, the strong form of functional stability is a special case of the weak form:*

$$\text{Strong form for } (f_i, d) \Leftrightarrow T_{\mathrm{rec}}^i(d) = 0.$$

*Proof.* (⇐) Suppose $T_{\mathrm{rec}}^i(d) = 0$. By Definition 4, for every $\varepsilon > 0$ there exists $\tau_\varepsilon < \varepsilon$ such that $Q_i^d(t) \ge Q_i^{\min}$ for every $t \ge t_0(d) + \tau_\varepsilon$. Hence $Q_i^d(t) \ge Q_i^{\min}$ for every $t > t_0(d)$. By continuity of $Q_i^d$ at $t = t_0(d)$, the same inequality holds at $t_0(d)$ itself, so the strong form holds.

(⇒) If the strong form holds, $Q_i^d(t) \ge Q_i^{\min}$ for every $t \ge t_0(d)$, so $\tau = 0$ satisfies the condition under the infimum, hence $T_{\mathrm{rec}}^i(d) = 0$. ▫

**Corollary 2.** *The parameter* $T^i_{\text{rec}}(d)$ *quantitatively measures the "distance" between the weak and strong forms for the pair* $(f_i, d)$.

*3.4.3. Relationships between parameters*

**Theorem 5 (Link between recovery time and degradation depth).** *For a system that satisfies the weak form,*

$$T^i_{\text{rec}}(d) > 0 \Leftrightarrow \Delta Q_i(d) > \delta_i.$$

*Proof.* By Theorem 4, $T^i_{\text{rec}}(d) = 0$ is equivalent to the strong form. By Theorem 2, the strong form is equivalent to $\Delta Q_i(d) \le \delta_i$. The contrapositive yields the claim. ▫

**Theorem 6 (Link between disturbance tolerance and admissible degradation).** *Fix a disturbance class k and define the worst-case response function*

$$\psi_i^{(k)}(a) = \sup_{t \ge t_0} \varphi_i^{(k)}(a, t).$$

*Then*

$$R_d^{i,k} = \sup\{a \in [0, A_k^{\max}]: \psi_i^{(k)}(a) \le \delta_i\}.$$

*Under continuity and strict monotonicity of* $\psi_i^{(k)}$ *on* $[0, A_k^{\max}]$, *the tolerance admits the closed-form expression*

$$R_d^{i,k} = \begin{cases} A_k^{\max}, & \psi_i^{(k)}(A_k^{\max}) \le \delta_i \text{ (class saturated within tolerance)}, \\ \left(\psi_i^{(k)}\right)^{-1}(\delta_i), & \psi_i^{(k)}(0) \le \delta_i < \psi_i^{(k)}(A_k^{\max}) \text{ (interior case)}, \\ 0, & \psi_i^{(k)}(0) > \delta_i \text{ (no admissible amplitude)}. \end{cases}$$

*Proof.* The strong stability condition $Q_i^{d_k(a)}(t) \ge Q_i^{\min}$ for every $t$ is equivalent to $\varphi_i^{(k)}(a, t) \le \delta_i$ for every $t$, which, taking the supremum over $t$, is equivalent to $\psi_i^{(k)}(a) \le \delta_i$. By Definition 3, $R_d^{i,k}$ is the supremum of amplitudes satisfying this condition, which yields the first equality. For the three cases: (i) if $\psi_i^{(k)}(A_k^{\max}) \le \delta_i$, the entire class is admissible; (ii) under strict monotonicity, $\psi_i^{(k)}$ has a continuous inverse on the interior range, and the equation $\psi_i^{(k)}(a) = \delta_i$ has a unique solution; (iii) if $\psi_i^{(k)}(0) > \delta_i$, no amplitude is admissible because the response at zero amplitude already exceeds the buffer. ▫

*Remark.* Under the standard zero-amplitude assumption $\varphi_i^{(k)}(0, t) = 0$ of Section 3.1, $\psi_i^{(k)}(0) = 0 \le \delta_i$ holds whenever $\delta_i \ge 0$ (i.e., $Q_i^{\min} \le Q_i^{\text{nom}}$), so case (iii) does not arise in the physical setting. The case is retained for formal completeness, since it captures configurations in which the nominal operating point itself sits below the admissible threshold (a misspecified design).

**Corollary 3.** *Disturbance tolerance* $R_d^{i,k}$ *and admissible degradation* $\delta_i$ *are related through the class-specific worst-case response* $\psi_i^{(k)}$. *An increase in* $\delta_i$ *(a* *wider admissible band) produces an increase in $R_d^{i,k}$ within every class, until saturation at $A_k^{\max}$. The separation of time and amplitude in the definition of $\psi_i^{(k)}$ allows the time-worst-case to be characterized independently of the amplitude argument.*

*3.4.4. Aggregate metrics and the forms of stability*

**Theorem 7 (One-directional link between strong stability and functional availability).** *If system $S$ satisfies the strong form of stability for $f_i$ over $D$, then for every observation horizon $T$ and every realized disturbance trajectory drawn from $D$, $A_i(T) = 1$.*

*Proof.* Under the strong form, $Q_i^d(t) \geq Q_i^{\min}$ for every $t \geq t_0(d)$ and every $d \in D$. The indicator $\mathbf{1}[Q_i(t) \geq Q_i^{\min}]$ is therefore 1 almost everywhere on the realized trajectory, so $A_i(T) = 1$. ▫

*Remark.* The converse implication requires the observation horizon to realize every class in $D$ at its worst admissible amplitude. Without that assumption, $A_i(T) = 1$ on a single trajectory only certifies the strong form on that trajectory and does not extend to the full set $D$. Analogously, $0 < A_i(T) < 1$ is consistent with the weak form when recovery occurs within the observation horizon. The case $A_i(T) = 0$ means that no admissible operation of $f_i$ is observed on the chosen horizon; it certifies violation of the weak form only if the horizon is known to exceed the admissible recovery budget or if the failed state persists indefinitely. Functional availability is therefore a *witness* of the stability form on observed trajectories rather than a complete equivalence with the worst-case definition over $D$.

**Theorem 8 (Properties of integral stability).** *The integral stability $\Psi(S, D) = \sum_{i=1}^{n} w_i \cdot \hat{R}_i \cdot T_i^{\max}/(T_i^{\max} + \overline{T}_{\text{rec}}^i)$ has the following properties:*

*(a) $\Psi(S, D) \in [0,1]$ for any system and any disturbance set.*

*(b) $\Psi = 1$ if and only if $\hat{R}_i = 1$ and $\overline{T}_{\text{rec}}^i = 0$ for every $i$ with $w_i > 0$, where these conditions are understood with respect to the specified disturbance-class sets $K_i$ and reference envelopes $R_{\text{ref}}^{i,k}$. The value $\Psi = 1$ certifies full normalized tolerance and zero recovery within the selected scope; it does not certify stability against disturbance classes outside $K_i$.*

*(c) For some function $f_i$ with $w_i > 0$, the mean recovery time $\overline{T}_{\text{rec}}^i \to \infty$ drives the contribution of that function to zero, so $\Psi$ decreases monotonically in $\overline{T}_{\text{rec}}^i$.*

*(d) $\Psi$ is non-decreasing in every $\hat{R}_i$ and non-increasing in every $\overline{T}_{\text{rec}}^i$.*

*Proof.* (a) Each factor $\hat{R}_i \in [0,1]$ and $T_i^{\max}/(T_i^{\max} + \overline{T}_{\text{rec}}^i) \in [0,1]$, so each summand is in $[0, w_i]$. Since $\sum_i w_i = 1$, $\Psi \in [0,1]$. (b) Each summand attains its maximum $w_i$ precisely when both factors are 1, i.e., $\hat{R}_i = 1$ and $\overline{T}_{\text{rec}}^i = 0$. (c) As $\overline{T}_{\text{rec}}^i \to \infty$, $T_i^{\max}/(T_i^{\max} + \overline{T}_{\text{rec}}^i) \to 0$. (d) Direct from the form of the summand. ▫

**Theorem 9 (Link between functional availability and the model parameters).** *Assume independent Poisson arrivals of the disturbance classes (class $k$ arriving at rate $\lambda_i^{(k)}$ at function $f_i$) and exponentially distributed repair (recovery) times. Consider a two-state failure–repair continuous-time Markov chain for function $f_i$, in which the function alternates between the working state $\{Q_i(t) \geq Q_i^{\min}\}$ and the failed state $\{Q_i(t) < Q_i^{\min}\}$. Let $p_i^{(k)}(R_d^{i,k})$ be the probability that a disturbance of class $k$ has amplitude exceeding $R_d^{i,k}$ (and therefore drives $f_i$ below the threshold). Let $\mu_i = 1/\overline{T}_{\text{rec}}^{i}$ be the recovery rate. Define the effective failure rate as*

$$\lambda_i^{\text{eff}} = \sum_k \lambda_i^{(k)} \cdot p_i^{(k)}(R_d^{i,k}).$$

*Then the steady-state functional availability is*

$$A_i = \frac{\mu_i}{\lambda_i^{\text{eff}} + \mu_i} = \frac{1/\overline{T}_{\text{rec}}^{i}}{\lambda_i^{\text{eff}} + 1/\overline{T}_{\text{rec}}^{i}}.$$

*Proof.* The two-state continuous-time Markov chain has transition rate $\lambda_i^{\text{eff}}$ from the working state to the failed state and $\mu_i$ from the failed state back to the working state. The balance equation $\pi_W \cdot \lambda_i^{\text{eff}} = \pi_F \cdot \mu_i$ together with $\pi_W + \pi_F = 1$ yields $\pi_W = \mu_i/(\lambda_i^{\text{eff}} + \mu_i)$. By ergodicity of the chain, $\pi_W$ equals the long-run fraction of time the function spends in the working state, which is $A_i$ by Definition 8. ▫

*Remark (degenerate case).* The CTMC formulation applies to the non-degenerate case $\overline{T}_{\text{rec}}^{i} > 0$. The zero-recovery case is handled by the limiting convention: as $\overline{T}_{\text{rec}}^{i} \to 0$, $\mu_i \to \infty$ and $A_i \to 1$, which is the correct value because either no threshold crossing occurs within the declared disturbance model, or the zero-recovery case is treated as the limiting strong-form regime. By the convention in Definition 9, $\overline{T}_{\text{rec}}^{i} = 0$ corresponds to the empty failure event, in which case $\lambda_i^{\text{eff}} = 0$ and the formula gives $A_i = 1$ directly.

**Corollary 4.** *Theorem 9 separates the contributions of disturbance tolerance and recovery time to availability. Disturbance tolerance acts through the probability term $p_i^{(k)}(R_d^{i,k})$: a higher per-class tolerance reduces the probability of threshold crossing and therefore the effective failure rate. Recovery time acts through the repair rate $\mu_i = 1/\overline{T}_{\text{rec}}^{i}$: faster recovery returns the function to the working state sooner. The two mechanisms combine in $A_i$ but enter distinct terms of the steady-state formula, which makes their independent effects identifiable from operational data.*

The structure of the connections between parameters and definitions is summarized in Figure 1.

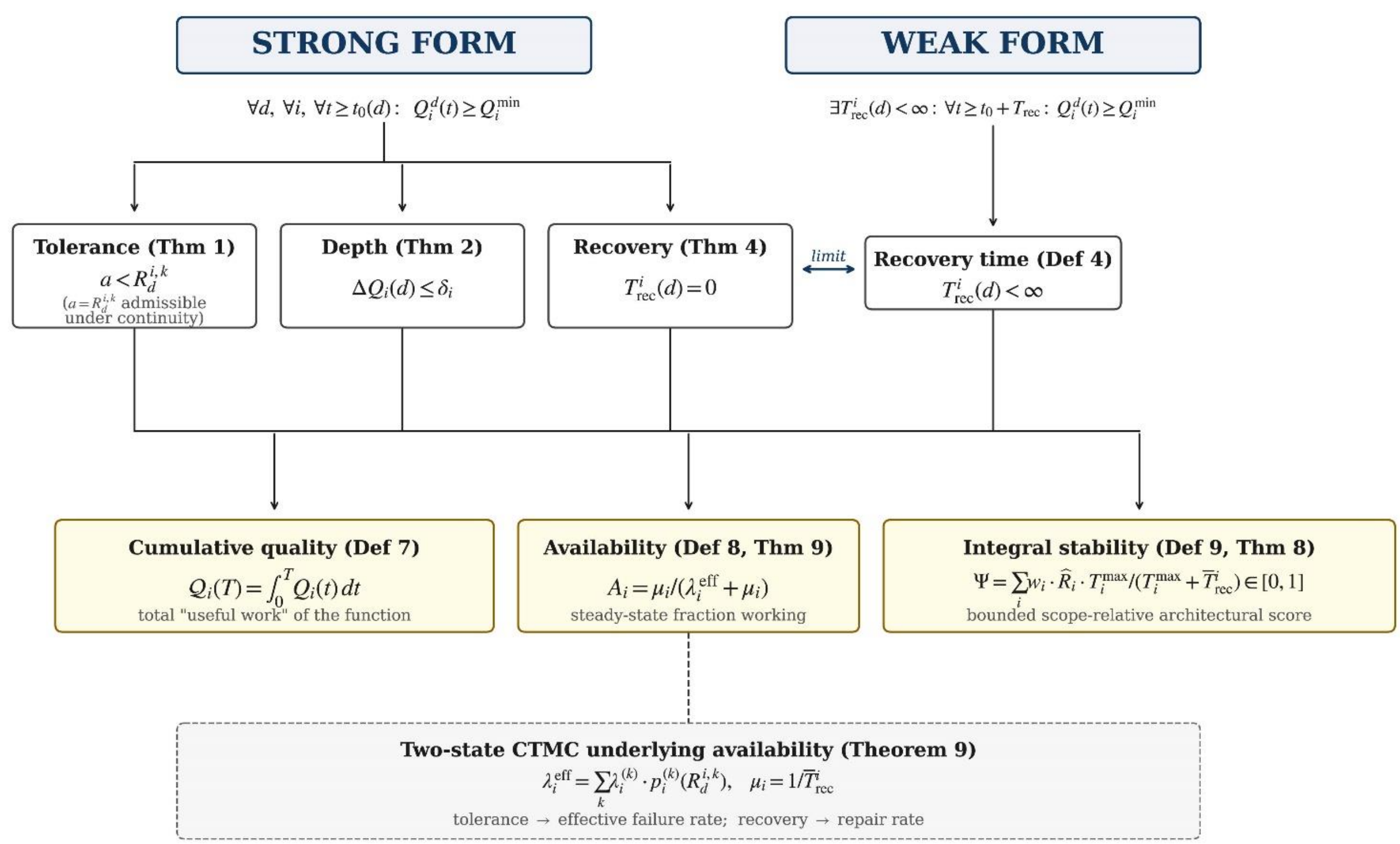

*Figure 1. Structure of the links between parameters and the formal definitions of functional stability. The strong form (top-left) admits three equivalent characterizations through tolerance (Theorem 1), degradation depth (Theorem 2) and zero recovery time (Theorem 4); the weak form (top-right) is characterized by a finite recovery time (Definition 4) and connects to the strong form through the* $T_{\text{rec}}^i \to 0$ *limit. The per-parameter view lifts to three aggregate metrics: cumulative quality* $Q_i(T)$*, functional availability* $A_i$*, and bounded integral stability* $\Psi \in [0,1]$*. Theorem 9 underlies the availability formula through a two-state failure–repair continuous-time Markov chain.*

# 4. Edge computing instantiation and case studies

## 4.1. Instantiation of the framework for edge computing

The abstract objects of Section 3 (functions, quality, disturbances, recovery) admit a direct mapping to the constructs of edge computing. The mapping is summarized in Table 2.

| Abstract object | Edge computing instantiation |
|---|---|
| System $S$ | An edge node, an edge cluster, or a segment of the edge–fog–cloud continuum |
| Function $f_i$ | A service deployed on the node: local inference, telemetry aggregation, control loop, video processing, model update |
| Quality $Q_i(t)$ | Utility-oriented per-service indicator: throughput, inference |

| Abstract object | Edge computing instantiation |
|---|---|
| | accuracy, frame rate, or a monotone transformation of latency such as $L_i^{\max}/L_i(t)$, energy budget, etc. |
| Threshold $Q_i^{\min}$ | Service-level objective at the node level: latency budget for a control loop, accuracy floor for an inference service |
| Disturbance set $D$ | Edge-specific disturbances structured by class $D_k$: backhaul-loss duration, CPU overload intensity, thermal-throttling factor, battery-depletion level, peer-node churn rate, DDoS amplitude, sensor degradation level, mobility velocity |
| Recovery time $T_{\text{rec}}^i(d)$ | Time to service migration to a neighbouring node, switch to a quantized local model, or re-offload to the cloud |
| Admissible degradation $\delta_i$ | Formal characterization of acceptable model quantization, frame-rate reduction, or accuracy-for-latency trade-off |
| Weight $w_i$ | Criticality of a service (a control loop carries higher weight than telemetry) |
| Integral stability Ψ | Bounded architectural figure of merit in [0,1] that aggregates per-service tolerance and recovery, weighted by criticality |

*Table 2. Mapping of the framework to edge computing constructs*

The mapping allows the formalism to absorb directly the patterns reported in the edge computing literature. Service migration [22, 23] manifests as a finite recovery time achieved by relocation of $f_i$ between nodes. Adaptive model selection [25, 38] manifests as operation in the band $[Q_i^{\min}, Q_i^{\text{nom}}]$ with bounded $\Delta Q_i$. Cloud fallback [48] manifests as a recovery mechanism whose latency budget is the dominant component of $T_{\text{rec}}^i$.

Below, the framework is applied to two analytical case studies that show how the metrics distinguish architectural alternatives that are equivalent under classical availability.

### 4.2. Case study 1: UAV swarm transporting a payload

Consider a system $S$ of four unmanned aerial vehicles (UAVs) operating as a swarm to transport a payload along a planned route. The swarm is treated as an edge cluster: each UAV is an edge node, and inter-UAV links form an opportunistic edge network. The set of functions is

- $f_1$: stabilization (per-UAV control loop);
- $f_2$: formation keeping (inter-UAV coordination);
- $f_3$: payload transport (collective lifting);
- $f_4$: telemetry to the ground station.

- Quality definitions:

– $Q_1$: control-loop frequency, with $Q_1^{\text{nom}} = 400$ Hz, $Q_1^{\text{min}} = 100$ Hz;

– $Q_2$: formation accuracy on the utility scale $Q_2(t) = 2 - |\text{err}(t)|$ for deviation $|\text{err}(t)| \le 2$ m, with $Q_2^{\text{nom}} = 2$ (zero deviation) and $Q_2^{\text{min}} = 0.5$ (1.5 m maximum admissible deviation);

– $Q_3$: lifting capability normalized by required capability; $Q_3^{\text{nom}} = 1.0$, $Q_3^{\text{min}} = 0.80$ (admissible payload deviation 20%);

– $Q_4$: telemetry update rate, with $Q_4^{\text{nom}} = 10$ Hz, $Q_4^{\text{min}} = 1$ Hz.

Criticality weights: $w_1 = 0.40$, $w_2 = 0.30$, $w_3 = 0.20$, $w_4 = 0.10$. Admissible recovery budgets: $T_1^{\text{max}} = 0.02$ s, $T_2^{\text{max}} = 1$ s, $T_3^{\text{max}} = 3$ s, $T_4^{\text{max}} = 1$ s.

The disturbance set $D$ is partitioned into classes $D_k$: loss of one UAV (motor failure or downing), inter-UAV link loss, wind gust parameterized by amplitude $a_{\text{wind}}$, and ground-link loss.

Two architectural alternatives are compared.

**Architecture A (no redundancy).** Each UAV carries a quarter of the payload, with no spare lift capacity and no peer takeover. Loss of one UAV reduces $Q_3$ from 1.0 to 0.75, *below* the threshold $Q_3^{\text{min}} = 0.80$, and recovery requires landing the payload and re-planning the mission, so $T_{\text{rec}}^3 = \infty$ for the in-flight scenario. Formation precision also degrades because the four-vertex configuration becomes a three-vertex configuration with $\hat{R}_2$ correspondingly reduced. The remaining functions on the surviving UAVs are unaffected: stabilization continues locally ($T_{\text{rec}}^1 = 0$), telemetry from the surviving UAVs continues at three-quarters of the nominal aggregate rate (7.5 Hz, above $Q_4^{\text{min}} = 1$ Hz, so $T_{\text{rec}}^4 = 0$).

**Architecture B (with redundancy).** Each UAV is designed with 35% lift capacity, so the four together provide 140% of the required payload, with 40% headroom. After loss of one UAV, the three remaining have 3 × 35% = 105% capacity, sufficient to carry the full payload. During the redistribution interval $Q_3$ dips to approximately 0.85, *above* the threshold $Q_3^{\text{min}} = 0.80$; the threshold is not crossed and therefore, by Definition 4, $T_{\text{rec}}^3 = 0$. The reconfiguration takes $T_{\text{nom}}^3 = 1.2$ s (return to the nominal 1.0 lift). Formation re-keeping also keeps $Q_2$ above its threshold, so $T_{\text{rec}}^2 = 0$ with $T_{\text{nom}}^2 = 0.8$ s. Local stabilization is unaffected on surviving UAVs ($T_{\text{rec}}^1 = 0$). Telemetry rate drops temporarily to 7.5 Hz, still above $Q_4^{\text{min}}$, so $T_{\text{rec}}^4 = 0$ with $T_{\text{nom}}^4 = 0.5$ s.

Diagnostic against the stability definitions:

– For $f_3$ in Architecture A: $\Delta Q_3 = 1.0 - 0.75 = 0.25 > \delta_3 = 0.20$, so the strong form is violated by Theorem 2. With $T_{\text{rec}}^3 = \infty$, the weak form is violated for the post-loss trajectory. On any observation horizon that starts after the UAV-loss event, $A_3(T) = 0$; on horizons that also include pre-loss operation, $A_3(T)$ equals the fraction of time before the loss and tends to zero as the post-loss horizon grows. The relevant normalized tolerance is $\hat{R}_3 = 0$.

– For $f_2$ in Architecture A: the three-vertex configuration produces persistent positional error that brings the formation accuracy near (but not below) the threshold; the design envelope of single-UAV loss is partially covered. Take $\hat{R}_2 = 0.6$ to reflect the reduced margin against wind disturbance, with $T_{\text{rec}}^2 = 0$ on nominal trajectories.
– For $f_3$ in Architecture B: $\Delta Q_3 = 0.15 < \delta_3 = 0.20$, so the strong form continues to hold for $f_3$ throughout the reconfiguration interval. $\hat{R}_3 = 1$ within the design envelope.
– For $f_1, f_2, f_4$ in Architecture B: all three functions stay above their thresholds under the design disturbance envelope, so $\hat{R}_i = 1$ and $T_{\text{rec}}^i = 0$. The reconfiguration durations are captured by $T_{\text{nom}}^i$.

Computing integral stability $\Psi$ from Definition 9 with the $\hat{R}_i$ values above and the recovery times derived in the diagnostic.

For Architecture A:

$$\Psi_A = 0.40 \cdot 1 \cdot \frac{0.02}{0.02 + 0} + 0.30 \cdot 0.6 \cdot \frac{1}{1 + 0} + 0.20 \cdot 0 + 0.10 \cdot 1 \cdot \frac{1}{1 + 0}$$

$$\Psi_A = 0.680.$$

For Architecture B:

$$\Psi_B = 0.40 \cdot 1 \cdot \frac{0.02}{0.02 + 0} + 0.30 \cdot 1 \cdot \frac{1}{1 + 0} + 0.20 \cdot 1 \cdot \frac{3}{3 + 0} + 0.10 \cdot 1 \cdot \frac{1}{1 + 0}$$

$$\Psi_B = 1.000.$$

Architecture B attains the maximum value $\Psi_B = 1.000$ within its design envelope, while Architecture A scores $\Psi_A = 0.680$. The 32% gap is driven by two distinct effects: the catastrophic failure of $f_3$ (the dominant component) and the reduced margin of $f_2$ (the secondary component). Classical system-level availability that averages over functions hides this gap; the per-function metric $\Psi$ exposes both effects directly. The reconfiguration durations $T_{\text{nom}}^i$ in Architecture B are not part of $\Psi$ because they do not represent threshold violations; they are reported as a separate quality-of-service indicator for transient performance.

### 4.3. Case study 2: edge AI inference with cloud fallback

Consider an edge AI inference service on a roadside unit (RSU) supporting vehicle-to-infrastructure (V2I) applications. Functions:

– $f_1$: local inference on the RSU using a quantized model;
– $f_2$: cloud-offloaded inference using a full-precision model;
– $f_3$: telemetry of inference quality back to the operator.
  - Quality definitions:
– $Q_1$: inference utility derived from end-to-end latency via $Q_1(t) = L_1^{\max}/L_1(t)$, where $L_1^{\max} = 100$ ms is the V2I latency budget; thus $Q_1^{\text{nom}} = L_1^{\max}/L_1^{\text{nom}} = 100/20 = 5$ and $Q_1^{\min} = 1$;

– $Q_2$: inference accuracy, with $Q_2^{\mathrm{nom}} = 0.96$ (full model), $Q_2^{\mathrm{min}} = 0.85$ (admissible for the application);
– $Q_3$: telemetry update rate, with $Q_3^{\mathrm{nom}} = 1$ Hz, $Q_3^{\mathrm{min}} = 0.1$ Hz.

Criticality weights $w_1 = 0.30$, $w_2 = 0.60$, $w_3 = 0.10$; admissible recovery budgets $T_1^{\max} = 0.2$ s, $T_2^{\max} = 0.5$ s, $T_3^{\max} = 2$ s.

The RSU normally offloads inference to the cloud when accuracy demand is high and falls back to the local quantized model when the uplink degrades. To keep the analytical comparison tractable, the numerical evaluation of $\Psi$ in this section restricts $K_2$ to the uplink-loss disturbance class; accuracy disturbances such as sensor noise and model drift are discussed qualitatively per architecture but are not included in the numerical value of $\Psi$. Three architectures are compared.

**Architecture I (cloud-only).** All inference is performed in the cloud. On uplink loss, $f_2$ fails; the service has no local fallback. $T_{\mathrm{rec}}^2 = T_{\mathrm{link}}$ (time to link restoration), which is unbounded for an outage of unknown duration. During outages, $Q_2 = 0 < Q_2^{\mathrm{min}}$, and the architecture violates the weak form when the link does not recover within the operational horizon. The normalized tolerance against the uplink-loss class is $\hat{R}_2 = 0$.

**Architecture II (edge-only).** All inference is performed locally on a quantized model with nominal accuracy approximately $0.87$. Under nominal conditions $Q_2 \geq Q_2^{\mathrm{min}} = 0.85$ holds, but the operating point sits only $0.02$ above the threshold. The uplink-loss class is irrelevant for this architecture (no link is used), so against $K_2 = \{\text{uplink-loss}\}$ the architecture is by construction unaffected. Qualitatively, accuracy disturbances such as sensor noise and model drift remain a concern: even small amplitudes push $Q_2$ below the threshold and in-deployment recovery requires offline retraining ($T_{\mathrm{rec}}^2 = \infty$ for that class). Restricting the numerical evaluation to the uplink-loss class yields $\hat{R}_2 = 1$ and $T_{\mathrm{rec}}^2 = 0$ for Architecture II; the resulting score should therefore be read as an upper bound that does not yet penalize accuracy disturbances.

**Architecture III (edge–cloud hybrid with fallback).** Inference is offloaded to the cloud when available; on uplink loss, the service switches to the local quantized model. The switch begins immediately on detection and the local model's accuracy of $0.87$ is reached after $50$ ms. During the transition window of approximately $50$ ms, $Q_2$ briefly drops below $Q_2^{\mathrm{min}}$; the threshold is crossed. By Definition 4, $T_{\mathrm{rec}}^2 = 50$ ms. After the switch, $Q_2 = 0.87 \geq Q_2^{\mathrm{min}}$; on link restoration, the service switches back to the cloud over a similar window. The weak form holds for $f_2$ globally with $T_{\mathrm{rec}}^2 = 50$ ms. Against $K_2 = \{\text{uplink-loss}\}$ the normalized tolerance is $\hat{R}_2 = 1$ (the architecture covers the full design envelope of single-link disturbances). For the considered accuracy disturbances, the cloud path can restore full-precision inference whenever the link is available, assuming that the underlying sensor data remain usable; therefore the operational tolerance is higher than for Architecture II alone, although a full accuracy-class analysis is left outside the numerical computation.

Computing integral stability $\Psi$ for the three architectures with $K_2 = \{\text{uplink-loss}\}$, $\hat{R}_1 = \hat{R}_3 = 1$, and per-function recovery times set as follows. Architecture I has $T^2_{\text{rec}} = \infty$ and $\hat{R}_2 = 0$. Architecture II has $T^2_{\text{rec}} = 0$ and $\hat{R}_2 = 1$ against the uplink-loss class (but is otherwise vulnerable, as discussed). Architecture III has $T^2_{\text{rec}} = 0.05$ s and $\hat{R}_2 = 1$. Other recovery times are taken as $T^1_{\text{rec}} = 0$ s, $T^3_{\text{rec}} = 0$ s.

$$\Psi_{\text{I}} = 0.30 \cdot 1 \cdot 1 + 0.60 \cdot 0 \cdot 0 + 0.10 \cdot 1 \cdot 1 = 0.40,$$
$$\Psi_{\text{II}} = 0.30 \cdot 1 \cdot 1 + 0.60 \cdot 1 \cdot 1 + 0.10 \cdot 1 \cdot 1 = 1.00,$$
$$\Psi_{\text{III}} = 0.30 \cdot 1 \cdot 1 + 0.60 \cdot 1 \cdot \frac{0.5}{0.5 + 0.05} + 0.10 \cdot 1 \cdot 1 = 0.945.$$

Against the uplink-loss class alone, Architecture II scores $\Psi_{\text{II}} = 1.00$ because it does not depend on the uplink, while Architecture III scores $\Psi_{\text{III}} = 0.945$ because of the brief transition window, and Architecture I scores $\Psi_{\text{I}} = 0.40$ because $f_2$ collapses. The ranking $\Psi_{\text{II}} > \Psi_{\text{III}} > \Psi_{\text{I}}$ correctly reflects robustness *to uplink loss alone* but is incomplete for the full design problem: Architecture II is unacceptable in deployment because accuracy disturbances bring $f_2$ below threshold without any in-deployment recovery.

To illustrate the scope-dependence, consider an illustrative full-envelope evaluation with $K_2 = \{\text{uplink-loss,accuracy-disturbance}\}$. For Architecture II the accuracy class binds: $\hat{R}_2 = 0.2$ (small drift margin) and $\overline{T}^2_{\text{rec}} = \infty$ (offline retraining), so the $f_2$ contribution collapses to zero and

$$\Psi^{\text{full}}_{\text{II}} = 0.30 + 0 + 0.10 = 0.40.$$

For Architecture III, assume the hybrid fallback maintains $\hat{R}_2 = 1$ across both classes (assuming that cloud inference compensates the considered accuracy disturbances whenever the link is available and the underlying sensor data remain usable) and $\overline{T}^2_{\text{rec}} = 0.05$ s for the combined envelope; then

$$\Psi^{\text{full}}_{\text{III}} = 0.30 + 0.545 + 0.10 = 0.945.$$

For Architecture I, $\hat{R}_2 = 0$ in both classes, $\Psi^{\text{full}}_{\text{I}} = 0.40$.

The two evaluations carry distinct meanings: edge-only is best when *only* uplink loss is considered, because it is by construction immune to that class; hybrid is best under the full operational envelope that includes both uplink loss and accuracy degradation. The framework does not “prefer” hybrid architectures; it honestly reports a ranking conditional on the declared scope. The Architecture III assumption $\hat{R}_2 = 1$ for the accuracy class is illustrative and would itself require justification in a real design study; the central methodological point is that ranking changes with $K_i$, and reporting $\Psi$ without an explicit $K_i$ is uninformative.

### 4.4. Summary of the case studies

Two regularities emerge from the case studies. First, the per-class disturbance tolerance $R^{i,k}_d$ and the per-function recovery time $T^i_{\text{rec}}$ separate architectures that share the same nominal availability. The classical availability coefficient, computed at the system level and aggregated over functions, is insensitive to the redistribution of operating margin between tolerance and recovery. Theorem 9 ties the two to availability through the two-state failure–repair model: tolerance enters through the effective failure rate $\lambda_i^{\text{eff}}$, while recovery time enters through the repair rate $\mu_i$.

Second, the per-function granularity reveals architectural defects that the system-level evaluation hides. Architecture A in the UAV case has acceptable system availability but $\hat{R}_3 = 0$ for the most critical lift function, which collapses the corresponding term in $\Psi$. Architecture I in the V2I case has acceptable nominal accuracy but fails the weak-form definition under sustained uplink outage. The V2I example also illustrates a third regularity: the reported $\Psi$ depends on the disturbance-class set $K_i$, and an incomplete $K_i$ can produce optimistic scores (Architecture II reaches $\Psi = 1.00$ against uplink-loss alone, although it is fragile against accuracy disturbances). Reporting $\Psi$ without an explicit $K_i$ is therefore not informative; comparison between architectures should fix the same $K_i$ across all candidates.

### 4.5. Simulation-based validation

To complement the analytical case studies of Sections 4.2–4.3, this subsection describes a reproducible simulation and testbed setup for empirical validation of the framework. The setup distinguishes two levels: a *simulation-based* level, on which the metrics $\hat{R}_d^{i,k}$, $T_{\text{rec}}^i$ and $\Psi$ are computed against controlled synthetic workloads, and a *testbed-based* level, on which the same metrics are measured on real hardware under injected disturbances.

*Simulation environment.* Two open-source tools cover the network and compute layers respectively. EdgeCloudSim [54] models edge–cloud placement, per-node CPU and memory saturation, and application-level metrics (task success rate, service time) directly. It is adapted here by adding two modules: a *disturbance injector* that draws events from a specified class $d \in D$ with amplitude sampled from $A_k(a)$, and a *quality recorder* that logs $Q_i^d(t)$ per function at a fixed sampling period. NS-3 [55] with the LTE and Wi-Fi modules covers the network-level disturbances (packet loss, latency spikes, backhaul link drops) at higher fidelity when the analysis needs correct queueing dynamics. The two tools are coupled through a shared event log so that a single disturbance sequence produces coherent traces on both. Running each architecture (A, B, I, II, III from Sections 4.2–4.3) under the same disturbance sequence and computing $\hat{R}_d^{i,k}$, $T_{\text{rec}}^i$ and $\Psi$ from the traces gives an *estimated* set of parameters that can be compared with the analytical predictions of Sections 4.2–4.3.

*Testbed setup.* A compact hardware testbed consists of three Raspberry Pi 5 boards (8 GB RAM each) connected through a managed Ethernet switch, running a k3s (lightweight Kubernetes) cluster with KubeEdge on top. Containerized services stand in for $f_1, \dots, f_n$: for the UAV case, a lift-control loop, a payload monitor and a telemetry publisher; for the V2I case, a YOLO-based inference service, a fusion service and a display service. Disturbances are injected using the Linux traffic control subsystem: `tc qdisc add dev eth0 root netem delay 200ms loss 5%` emulates link degradation between two nodes; `stress-ng --cpu 4 --cpu-load 80` induces CPU throttling; a scripted event generator triggers container evictions and pod restarts through the Kubernetes API. The Prometheus stack collects per-function metrics; a small analysis script computes $\hat{R}_d^{i,k}$ and $T_{\mathrm{rec}}^i$ from the recorded time series and reports $\Psi$ per architecture. The setup is small enough to be reproduced by a reader in a single afternoon.

*Expected outputs.* Under the disturbance class set $K_i$ declared in each case study, the simulation and the testbed should yield estimated $\hat{R}_d^{i,k}$ within one order of magnitude of the analytical values (accounting for finite-sample noise) and $T_{\mathrm{rec}}^i$ within a factor of two of the analytical bounds. The ordering of architectures by $\Psi$ should agree with the analytical ordering in Sections 4.2 and 4.3: $B > A$ for the UAV case (with the same $K_i$), $II > III > I$ for the V2I case restricted to uplink-loss disturbances, and $III > II \approx I$ when accuracy disturbances are also included in $K_i$. The specific numerical values depend on the disturbance amplitude scaling and are not fixed by the framework itself; the criterion of empirical validity is qualitative agreement of the orderings and quantitative agreement within these tolerance bands.

A full simulation and testbed campaign is beyond the scope of the present article, whose contribution is the formal framework and its analytical instantiation. The setup described above is a concrete plan for the follow-up empirical study, giving reviewers and prospective users a self-contained recipe rather than a promise.

# 5. Discussion

## 5.1. Comparison with existing concepts

The detailed conceptual comparison between functional stability and the three established concepts (resilience, fault tolerance, dependability) is summarized in Table 3.

| Criterion | Resilience [8] | Fault tolerance [9] | Dependability [10] | Functional stability |
|---|---|---|---|---|
| Focus | Process (resistance + recovery) | Continuation under faults | Set of attributes | Per-function quality |
| Unit of analysis | System | System under faults | System as product | Individual functions |
| Metric | Binary or temporal | Binary | Set of attributes | Quantitative ($Q_i \geq Q_i^{\min}$) |
| Degradation | Implicit | Not modelled | Not specified | Explicit |

| Criterion | Resilience [8] | Fault tolerance [9] | Dependability [10] | Functional stability |
|---|---|---|---|---|
| | | | | ($\delta_i$) |
| Granularity | System | System | System | Per-function |
| Time dynamics | Recovery | Continuity | Static | $T_{\text{rec}}^i$ per function |
| Disturbance model | External | Internal faults | Implicit | Explicit set $D$ |

*Table 3. Comparison of system stability concepts*

The position of functional stability among these concepts is complementary rather than competitive. The strong form connects directly to the zero-downtime ideal of high-end fault tolerance and to the resilience concept in its strictest interpretation. The weak form formalizes graceful degradation as a first-class property rather than as an auxiliary mechanism. The integral stability $\Psi \in [0,1]$ provides a single bounded figure of merit that is sensitive to the same properties (recovery time, operating margin, criticality) that operations teams already track separately through MTTR, error budgets and SLO attainment in SRE practice [50].

### 5.2. Relation to edge computing practice

The framework aligns with three established edge computing practices.

*Service-level objectives.* The per-function threshold $Q_i^{\min}$ is, in operational terms, the SLO for the service $f_i$ on the edge node. The framework lifts SLOs from an operational construct to a formal one: the strong form is the condition that SLOs are continuously satisfied; the weak form is the condition that SLOs are restored within a bounded time after any disturbance in $D$.

*Service migration.* Decisions about whether and when to migrate a service between edge nodes or between edge and cloud are typically optimized against local objectives [22, 23]. The integral stability $\Psi$ provides a system-level objective against which migration policies can be evaluated. A policy that reduces $T_{\text{rec}}^i$ for critical services raises $\Psi$; one that reduces it for low-weight services has limited effect.

*Adaptive computation.* Switching between full and quantized models, between local and offloaded inference, and between full frame rate and reduced frame rate corresponds to operation in the band $[Q_i^{\min}, Q_i^{\text{nom}}]$. The framework formalizes this band through the parameter $\delta_i$ and makes explicit the requirement that the band has positive width: an SLO set too close to the nominal operating point leaves no room for adaptive computation and forces the system into the strict regime of the strong form.

*Integration with edge orchestrators.* The integral metric $\Psi$ can be embedded as a decision signal in edge orchestration platforms such as Kubernetes and KubeEdge.

A concrete integration path is as follows. Per-function quality trajectories $Q_i^d(t)$ are collected through the standard Prometheus/OpenTelemetry pipeline already available in Kubernetes deployments, using service-specific exporters that emit latency, accuracy drift, throughput shortfall and frame-rate deficit as time-series metrics. A sidecar component computes running estimates of $\hat{R}_d^{i,k}$, $T_{\text{rec}}^i$ and $\Psi$ over a rolling window from these series. The estimated $\Psi$ is then exposed either as a custom Horizontal Pod Autoscaler (HPA) metric to trigger scale-out when $\Psi$ falls below a threshold, or as a scheduler priority signal through the Kubernetes scheduler framework to bias pod placement toward nodes whose $\Psi$-contribution is highest for the pod's service class. On KubeEdge, where node-cloud communication is intermittent, the same signal drives the *edgemesh* traffic policy and the migration decisions made by the *cloud-edge* controller: a pod whose function has $\Psi$-critical weight $\alpha_i$ and observed $T_{\text{rec}}^i$ approaching the SLO ceiling becomes a migration candidate, while pods with low $\alpha_i$ tolerate longer recovery periods on the current node. This turns the framework into an operational tool for site reliability engineering (SRE) teams: $\Psi$ provides a single figure of merit that the scheduler and the SRE dashboard both consume, replacing the ad-hoc composition of per-service SLO alarms with a system-level objective grounded in the theorems of Section 3.4.

### 5.3. Limitations

The framework has three limitations that point to directions for further work.

First, the framework treats the function decomposition $F = \{f_1, \dots, f_n\}$ as given. In practice, the decomposition is itself a design decision: choosing the granularity at which services are defined affects every parameter of the framework. The relationship between decomposition granularity and functional stability is an open question that connects to the architectural literature on microservice granularity [51].

Second, the disturbance set $D$ is treated as a structured but otherwise external input. In edge environments, disturbances are partially endogenous: a heavy inference request can itself produce thermal throttling that disturbs other services on the same node. A more complete framework would model such feedback loops, possibly through compositional reasoning of the kind developed for hybrid automata [52].

Third, the case studies in Section 4 are analytical, and the simulation and testbed setup described in Section 4.5 is a plan, not a completed empirical campaign. The present article contributes the formal framework, its parameters and their linking theorems, together with a reproducible experimental protocol. A follow-up empirical study running the described EdgeCloudSim/NS-3 simulation and the k3s/KubeEdge testbed under the disturbance-class set of each case study, and reporting the estimated $\hat{R}_d^{i,k}$, $T_{\text{rec}}^i$ and $\Psi$ against the analytical predictions, is the natural continuation of the cycle.

## 6. Conclusions and future directions

This article proposes a formal framework of functional stability for edge computing systems. The framework shifts the unit of analysis from the system to the individual function and replaces the binary working/failed evaluation with a continuous quality function. The two forms of the definition (Definitions 1 and 2) cover both zero-downtime systems and graceful-degradation systems. A system of base parameters and aggregate metrics (Definitions 3–9) is connected to the formal definitions by nine theorems (Theorems 1–9). The framework is instantiated for edge computing through a mapping table (Table 2) and illustrated by two case studies that show both how the metrics distinguish architectural alternatives equivalent under classical availability and why the disturbance-class set must be fixed explicitly before comparing designs.

The principal contributions are: (i) the per-function formalization of stability with continuous quality and per-function thresholds; (ii) the system of base and aggregate parameters with explicit theorems linking them to the definitions; (iii) the edge computing instantiation and the demonstration on two analytical case studies that the integral stability metric is sensitive to architectural choices that classical availability does not separate, while also exposing the dependence of rankings on the declared disturbance-class scope.

Directions for further work follow from the limitations identified in Section 5.3. The relationship between function decomposition granularity and stability parameters is an open architectural problem. Endogenous disturbance modelling, in which disturbances are partly produced by the services themselves, calls for compositional reasoning developed for hybrid automata [52]. Empirical validation on the simulation and testbed setup described in Section 4.5 (EdgeCloudSim/NS-3 simulation coupled with a three-node k3s/KubeEdge cluster on Raspberry Pi hardware under `tc/netem` and `stress-ng` disturbance injection) is the next concrete step. Integration of the integral metric $\Psi$ as a scheduling signal in Kubernetes and KubeEdge orchestrators (Section 5.2) transforms the framework from a formal artefact into an operational tool for site reliability engineering. Stochastic generalizations of the framework, in which disturbances are characterized by distributions and the strong and weak forms are interpreted in probabilistic terms, are a natural extension. Automated verification of the functional stability conditions through model checking [53] would close the loop from formalization to engineering practice.

## Acknowledgements

The author thanks the colleagues at the Department of Software Systems and Technologies of the Faculty of Information Technologies, Taras Shevchenko National University of Kyiv, for discussions that shaped the presentation.